\documentclass[final,3p,times,twocolumn]{elsarticle}
\usepackage{hyperref} 
\usepackage{wrapfig}
\usepackage{colortbl}
\usepackage{caption}
\usepackage{subcaption}
\usepackage{amssymb}
\usepackage{amsfonts}
\usepackage{enumitem}
\usepackage{algorithm}
\usepackage{algpseudocode}
\usepackage{color}
\usepackage{xcolor}
\usepackage{xspace}
\usepackage{courier}
\usepackage{xpatch}
\usepackage{listings}
\usepackage{realboxes}
\usepackage{array}

\usepackage{mathtools}

\hypersetup{
  colorlinks=false,
  linkcolor=darkred,
  citecolor=darkgreen,
  urlcolor=darkblue
}

\usepackage[noabbrev,capitalise]{cleveref}

\newcounter{ex}

\newtheorem{example}[ex]{Example}

\usepackage[utf8]{inputenc}

\usepackage{tikz}
\usetikzlibrary{backgrounds, decorations.pathreplacing, positioning,calc,shapes,arrows,positioning}

\usepackage{listings}

\lstdefinestyle{scode}{
language=C++,
mathescape, 
numbers=left,  
escapechar = !, 
directivestyle={\color{black}},
columns=flexible,
commentstyle=\color[rgb]{0.5,0.5,0.5},  
}

\definecolor{gray}{gray}{0.90}
\definecolor{lightblue}{RGB}{173, 216, 230}
\definecolor{lightred}{RGB}{255, 200, 200} 
\definecolor{blue}{HTML}{0066cc}
\definecolor{purple}{HTML}{660099}

\newcommand{\red}[1]{\textcolor{red}{#1}}
\newcommand{\purple}[1]{\textcolor{purple}{#1}}

\newcommand{\orange}[1]{\textcolor{orange}{#1}}

\newcommand{\blue}[1]{\textcolor{blue}{#1}}

\newcommand{\code}[1]{{\small\texttt{#1}}}

\newcommand{\stitle}[1]{\smallskip\noindent\textbf{#1\xspace}}

\usepackage{amssymb}
\usepackage{amsmath}

\journal{VSI: HILDA}

\begin{document}

\begin{frontmatter}



\title{A Decade of Systems for Human Data Interaction}


\author[aff1]{Eugene Wu}
\author[aff1,aff2]{Yiru Chen}
\author[aff1]{Haneen Mohammed}
\author[aff1,aff3]{Zezhou Huang}

\address[aff1]{Columbia University}
\address[aff2]{Adobe}
\address[aff3]{Microsoft}

\begin{abstract}
Human–data interaction (HDI) presents fundamentally different challenges from traditional data management. HDI systems must meet latency, correctness, and consistency needs that stem from usability rather than query semantics; failing to meet these expectations breaks the user experience. Moreover, interfaces and systems are tightly coupled; neither can easily be optimized in isolation, and effective solutions demand their co-design. This dependence also presents a research opportunity: rather than adapt systems to interface demands, systems innovations and database theory can also inspire new interaction and visualization designs. We survey a decade of our lab's work that embraces this coupling and argue that HDI systems are the foundation for reliable, interactive, AI-driven applications.



\end{abstract}


\begin{keyword}
Data Visualization,
Database Theory,
Human Data Interaction



\end{keyword}

\end{frontmatter}


\section{Introduction}

Human–Data Interaction (HDI) studies how people access, understand, manipulate, and manage data. It spans scenarios such as querying databases through natural language interfaces~\cite{li2014constructing}, performing large-scale visual analytics~\cite{heer2010zoo}, and managing personal or organizational data~\cite{mortier2014human}.
Among these settings, data visualization interfaces are especially demanding: they must communicate complex information effectively {\it and} support expressive, low-latency interactions at the pace of human thought. As a result, visualization interfaces push the limits of both data systems and user interface design.

A key challenge is that HDI spans two traditionally independent research traditions.
Data management approaches HDI by designing data-centric systems and abstractions.
As application demands have evolved, so have the underlying data systems: transactional workloads motivated ACID semantics and concurrency control; analytical workloads motivated columnar storage and physical database design; web-scale applications motivated distributed data stores and relaxed consistency models; and streaming workloads motivated incremental view maintenance and real-time processing.
Throughout these shifts, data management research takes a holistic view that co-evolves physical design, query process, system architecture, and developer abstractions in response to new data needs.


Visualization, in contrast, starts with the user. As Card et al. write, it focuses on  {\it ``interactive, visual representations of data to amplify cognition~\cite{card1999readings},''}  while Heer and Kandel emphasize the need to support  \textit{exploration at the rate of human thought~\cite{heer2012interactive}.''} 
These definitions highlight how visual representation empowers users, but remain agnostic to critical system concerns---such as how data is modeled, how analyses are executed, or how the system is architected and abstracted---which are central to data management research.

Historically, the separation was effective. Visualization tools focus on user experience, while DBMSes optimize performance and scalability. Small datasets can be run in the browser~\cite{heer2006software, bostock2011d3, satyanarayan2017vega}, while larger workloads simply send SQL to a backend and rely on the DBMS's performance. Tools like Tableau and Looker embody this loose coupling, where the client translates user interactions into queries sent to backend data systems.


Increasingly, this traditional separation is breaking down because visualizations fundamentally abstract away data scale: a single visual mark may summarize one record or a billion, yet the user's responsiveness expectations remain the same. 
This abstraction hides the true computational cost of interactions, while the interface continues to generate rapid, fine-grained updates (e.g., brushing, sliders, coordinated filters). As a result, the underlying system must satisfy interactive latency guarantees despite huge variations in data sizes and query complexity.  This is difficult to achieve in a one-size-fits-all manner, and meeting those guarantees requires tight coupling between the visual design, its interaction semantics, and the physical execution strategies beneath it. Classic DBMS optimizations alone are insufficient because the interface’s semantics, structure, and interaction model now directly determine what system designs are feasible at interactive timescales.



\smallskip
\noindent We refer to this coupling as \textbf{Design Dependence}, where

\begin{itemize}[leftmargin=*, itemsep=0em]
    \item the {\it Interface Design} imposes latency, interaction semantics, and consistency demands on the system that necessitate specialized architectural designs, and
    \item the {\it System Design} constrains what interactions are feasible at interactive timescales.
\end{itemize}

\begin{figure}
    \centering
    \includegraphics[width=.35\linewidth]{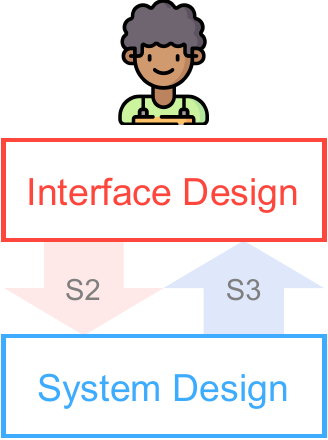}
    \caption{Design dependence describes how interface designs necessitate system complexity, but system designs also limit the capabilities of the interface.   This dependence necessitates a holistic approach to human data interface design.  The dependencies are labeled with the sections that describe them.}
    \label{fig:overview}
\end{figure}

\noindent This {\it design dependency} (Figure~\ref{fig:overview}) goes beyond performance because new system abstractions can enable new classes of interaction design, while new visual designs or interaction requirements motivate new system designs. For example, streaming databases like Materialize enable real-time maintenance of predefined views but, in turn, constrain dashboards to interactions that efficiently map to those maintained views~\cite{mcsherry2013differential}. In short, system design shapes interaction design, and interaction design shapes system design .

As data scale, interface complexity, and user expectations continue to rise, HDI systems cannot be built through isolated optimizations or point solutions. Designers need the ability to co-design interfaces and systems, and future HDI research should focus on abstractions, frameworks, and tools that accelerate and optimize this iteration cycle.
\textbf{Supporting this co-evolution is the central systems challenge for the next generation of HDI research.}


Finally, Design Dependence also affects broader research directions: visualization research places demands on systems, and systems theory often reveals new opportunities for interaction and representation. Understanding this virtuous cycle is essential, especially as emerging AI technologies blur the boundary between interface and system.

In the sections that follow, we illustrate this thesis using a decade of research from our lab. \Cref{s:interface} shows how formalizing interface behavior can drive system co-design and optimization. \Cref{s:system} shows how systems theory and data models can inspire new interaction techniques and visualization theory. We conclude with a perspective on AI’s role in the future of HDI systems.

\stitle{Terminology:} We use the terms {\it visualization} or {\it view} to refer to rendered visual output (e.g., charts, maps); {\it interface} to refer to the user-facing front end (layout, views, interactions); and {\it system} to refer to the underlying data-processing architecture (e.g., DBMS, execution engines, communication) that manages resources and processes data.

\section{From Interface to System Design}\label{s:interface}

Interfaces introduce analytical patterns and interaction demands that exceed what existing systems are designed to support.
This raises a core systems question: given an interface $I$ and limited engineering resources, how do we identify the necessary system optimizations and the acceptable performance–resource trade-offs?

\begin{figure}
    \centering
    \includegraphics[width=0.9\linewidth]{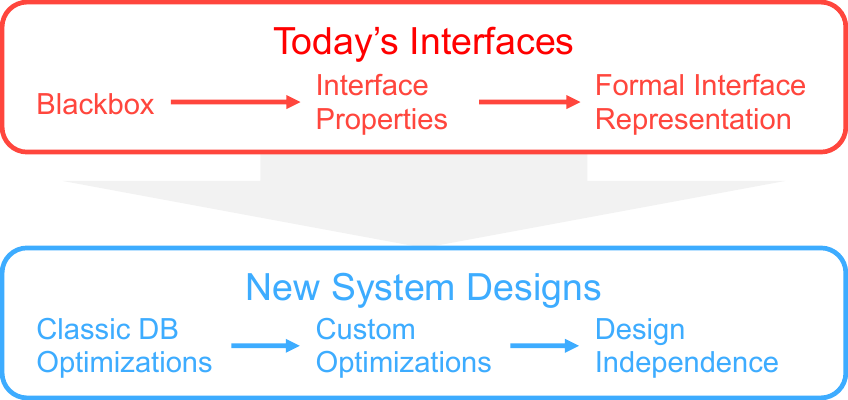}
    \caption{Interface designs necessitate new system designs to meet analysis needs and interactivity expectations.  Systems are increasingly shifting from a black box model of the interface, to one that leverages specific interface properties, to one that leverages a formal representation of the interface. }
    \label{fig:i2s}
\end{figure}

This section introduces a design spectrum for how systems can characterize $I$ to design new optimizations (Figure \ref{fig:i2s}). The spectrum has three points:
\begin{itemize}[leftmargin=*,itemsep=0em]
    \item \stitle{Black-box.}   At one extreme, the interface is treated as a black-box workload
$I\sim Q=\{q_1,\dots\}$.  This fully decouples interface and system, but prevents the system from exploiting any interface-level semantics.
    
    \item \stitle{Interface Properties.} In the middle,
    $I=(Q, p_1,\dots,p_k)$ 
    augments the workload with a set of interface properties $p_i$ that the system can safely assume and aggressively optimize for. This approach is powerful but brittle: developers must manually identify these properties and select optimizations that exploit them, a task that becomes increasingly untenable as new optimizations proliferate and the optimization space grows combinatorially.

    \item \stitle{Formal interface representation.} At the other extreme, a formal representation of the interface, user, and task could enable holistic analysis and direct system optimization.
    However, designing a representation that is both expressive and amenable to optimization remains an open research question.
\end{itemize}


\subsection{Black-box Query Workload}

A large body of prior work treats the interface as a black box that produces a stream of queries.
This model is attractive because it mirrors how most visualization systems operate today: user interactions (e.g., brushing, filtering, or zooming) are translated into SQL queries that the underlying DBMS executes to update visible views.
Under this model, the interface reduces to a traditional analytical workload, and  decades of database research that reduce query latency and improve throughput should, in principle, directly benefit interactivity\footnote{For a comprehensive review, see Battle et al.~\cite{battle2020structured}.}.   This model also makes it straightforward to create interaction benchmarks~\cite{eichmann2020idebench,battle20benchmark}.

Much of the systems literature on visualization performance adopts this viewpoint.
For example, data cubes reduce the cost of common aggregation queries by pre-computing summaries over low-dimensional group-bys that are often small enough to cache in memory~\cite{liu2013immens,lins2013nanocubes,pahins2016hashedcubes}.
Similarly, indexes accelerate pan-and-zoom interactions and nearest neighbor look ups in map-based or spatial interfaces~\cite{kyrix}. 
Because these structures align with familiar analytical patterns, they can be implemented directly on existing DBMSs or embedded engines. Mosaic exemplifies this approach by building partial data cubes via SQL and evaluating them client-side using an embedded DBMS~\cite{mosaic} rather than relying on bespoke data structures.

Unfortunately, simply reducing query latency does not guarantee better interactivity.
Standard metrics such as p95 latency often fail to correlate with perceived responsiveness~\cite{jiang2018evaluating,rahman2020evaluating}, and reflect a mismatch between system-level performance indicators and user experience.
Treating the interface purely as a query workload source abstracts away critical aspects of interaction design, perception, and expectation. 
For example, a nominally simple action—like adjusting a slider—may trigger a burst of concurrent queries, many of which update off-screen or low-priority views. Executing all queries at once forces them to compete for shared resources and can delay the updates the user is actively attending to.

Prior black-box approaches have demonstrated that conventional database optimizations can substantially improve performance, but also reveal their limitations. This model abstracts away the interaction semantics, perceptual constraints, and task structure that ultimately determine whether users experience the system as fluid. These limitations motivate approaches that incorporate additional interface properties or, as we argue in later sections, richer representations of the interface itself.

\subsection{Interface Properties} 

As the limitations of the black-box workload model suggest, optimizing standard workload metrics does not always lead to fluid user experiences~\cite{jiang2018evaluating}. Modern data interfaces can easily generate enough queries to saturate compute, memory, and/or network resources. Thus, the challenge is not to simply execute every query quickly, but to allocate resources to the queries that best meet the user's current needs.

A substantial body of prior work addresses this gap by exploiting interface properties---assumptions about the query, the visual representation, the interaction design, the task, or the user’s behavior that the system can safely rely on.   Such examples include:
\begin{itemize}[leftmargin=*, itemsep=0em]
    \item user prefers fast \& coarse over slow \& exact answers,
    \item user is on a mobile device and is bandwidth-limited,
    \item query is a group-by aggregation without joins,
    \item user is unlikely to revisit portions of the application,
    \item or user will not detect certain visual differences.
\end{itemize}


These properties help systems implement aggressive optimizations that would not apply in general settings.
Below, we illustrate this idea using two common assumptions---network bottlenecks and perceptual tolerance---to highlight the power and weaknesses of relying on interface properties.  

\subsubsection{Network Communication}
Many interactive data interfaces are deployed as web-based client–server applications, where network latency and bandwidth often dominate response time. Prior work exploits this property to optimize the client-server communication. Classic approaches such as prefetching predict what the user is likely to do next and proactively send responses before the user's request is issued. This is effective when requests are sparse and predictable, such as in document retrieval or web search, where task models and behavioral signals provide reliable cues and there is enough time between the prediction and user's request to perform the prefetch~\cite{battle2016dynamic}.




Visualization interactions, however, exhibit very different characteristics. Continuous interactions such as brushing or sliders can generate dozens of requests per second, which leaves little opportunity for prefetching.  Even when the predictor is perfect, the server still must wait for each prefetch request to arrive before sending a response, by which point the user may have already moved on. The standard assumptions behind web prefetching (that full responses are more important than latency, and that requests are sparse) do not hold.


Motivated by these constraints, our Khameleon~\cite{mohammed2020continuous} system reframes the problem: it guarantees immediate responses and treats fidelity as the optimization objective. Khameleon progressively encodes results so that any prefix can be rendered, giving the client flexibility to hedge across many possible future interactions or to store full-fidelity results for a few likely ones. Client messages (including explicit prefetch requests) serve only as hints that help the server allocate its network budget more effectively; the server proactively pushes partial results and remains decoupled from the client.



We found that even with a perfect predictor of the next five requests, prefetching on an LTE network exhibited median $50{s}$ latencies, and ${<}30{\%}$ of requests completed within $10{s}$. Khameleon reduced latency to $30{ms}$---a $1,666\times$ improvement---and converged to full fidelity within $1{s}$. Of course, this requires additional server resources to perform scheduling and designing efficient encoding schemes.



\subsubsection{Approximation}
The second property is to leverage imperfect user senses. Decades of work in perception and psychophysics, such as Weber’s just-noticeable difference (JND)~\cite{jnd,fechner1860elemente}, seek to quantify the extent humans can detect numerical differences. These insights motivated the rich literature on approximate query processing (AQP)~\cite{aqp,agarwal2013blinkdb,hellerstein1997online}, and more recently, on approximations targeted specifically at visualization.

Visualization-oriented approximation typically seeks to preserve visual rather than statistical fidelity. Because the final visualization has finite pixel resolution, techniques such as M4~\cite{m4} and our work on OM3~\cite{om3} exploit ``resolution push-down'' to pre-aggregate statistics at the pixel level so that the visualization can be reconstructed exactly from coarse summaries.  This shifts the cost from network transfer to database computation. OM3 further provides progressive prefixes that render increasingly refined versions of the chart, mirroring Khameleon’s progressive encoding strategy.



When approximations diverge from exact results, systems must assume the types of errors users find acceptable and find ways to convey the uncertainty. However, users struggle to interpret statistical uncertainty~\cite{Hullman2019Why,Padilla2021Uncertainty,Kahneman1982Judgment}.  Further, accuracy guarantees over individual aggregates are not enough, because they may not preserve higher-level patterns that guide visual interpretation. 
 BlinkDB~\cite{agarwal2013blinkdb} addresses these issues by exposing user-specified error or time budgets, but users find it hard to set these parameters.   
Distribution-oriented methods seek to preserve shapes or density profiles~\cite{sampleseek,macke2021rapid}, or to preserve properties of functions over the resulting visualization; for instance, Rahman et al.~\cite{rahman2017ve} preserves pairwise orderings in bar charts.


Unfortunately, perceptual error tolerances vary across people, tasks, and encodings and are not yet fully characterized. This makes property-driven approximations powerful but brittle. 
An open question is how to define perceptual models that are both general enough to apply across visualizations and precise enough to support optimization. Such models would allow systems to reason about approximation error in terms of perceived fidelity rather than statistical metrics. 
For instance, our work on perceptual functions (p-funks)~\cite{alabi2016pfunk} represents an early attempt to formalize these constraints.  It specifies a just-noticeable difference (JND) model as function of mark properties (e.g., pixel height) so the system can guarantee that online approximations remain within perceptual bounds. This illustrates how interface-aware perceptual models can guide system behavior, but also highlights the need for more principled, general representations of visualization semantics.



These communication and approximation examples illustrate the potential and the limitations of property-driven optimization. Interface properties allow systems to exploit assumptions about user behavior, visual perception, and interaction structure that would be impossible under a general analytical workload. Yet each optimization depends on a narrow set of assumptions holding true. As interfaces grow more complex and analyses span multiple coordinated views, interaction types, and data sources, relying on manually identified properties becomes increasingly brittle and difficult to scale.

\subsubsection{A Deluge of Special Cases}

Interface designers face a difficult and increasingly untenable position. 
To achieve interactive performance, they must choose layouts, encodings, and interaction patterns that align with a sprawling ecosystem of bespoke system optimizations---each tailored to a narrow performance sweet spot under specific interface or user assumptions. Simply understanding the subtle distinctions between dozens of approximation, caching, and scheduling strategies is challenging.  Further deciding which combinations apply to a particular interface is beyond the reach of most developers.   

Worse, these optimizations rarely cleanly compose. Techniques that preserve visual salience may conflict with those that guarantee ordinal consistency, and both may be incompatible with progressive scheduling strategies such as those used by Khameleon. \textbf{As researchers develop new optimizations, the burden on designers and developers only increases.} 
The interface-system design space grows combinatorially with the number of views, interactions, analytical tasks, and database complexity, making manual enumeration and special-case engineering impractical outside of narrowly crafted settings.

We need a principled interface representation to guide system-level decisions---akin to how a query optimizer uses a logical plan to reason about physical execution. Without such representations, developers are forced to stitch together one-off optimizations, unable to generalize or automate across interfaces.

\subsection{Formal Representations \& Design Independence}\label{ss:pvd}

The brittle nature of special-case optimizations highlights the need for \textbf{Design Independence}: a principled way to represent interface semantics so the system, rather than the designer, can decide how to meet interactivity constraints. Design Independence would let interface designers  rapidly iterate without being bottlenecked by checking with developers or architects about implementation feasibility~\cite{Walny2019DataCE}, much as logical plans let query optimizers reason about execution without exposing physical details.

\subsubsection{Physical Visualization Design}

 Physical Visualization Design (PVD)~\cite{chen2025physical} is our approach towards Design Independence. Similar to physical database design (PDD)~\cite{christian2011autoadmin,zilio2004db2,finkelstein1988physical}, which selects physical structures and layouts to accelerate a representative query workload, PVD selects data structures, caching strategies, and execution plans to meet the latency constraints of a visualization interface. 
The key idea is to provide a formal yet lightweight representation of the interface that captures the data flow, interactions, and latency requirements.  This representation is input to an automated optimizer can analyze and synthesize a feasible client–server–DBMS execution strategy, and then instantiate the architecture.


\begin{figure}
    \centering
    \includegraphics[width=\linewidth]{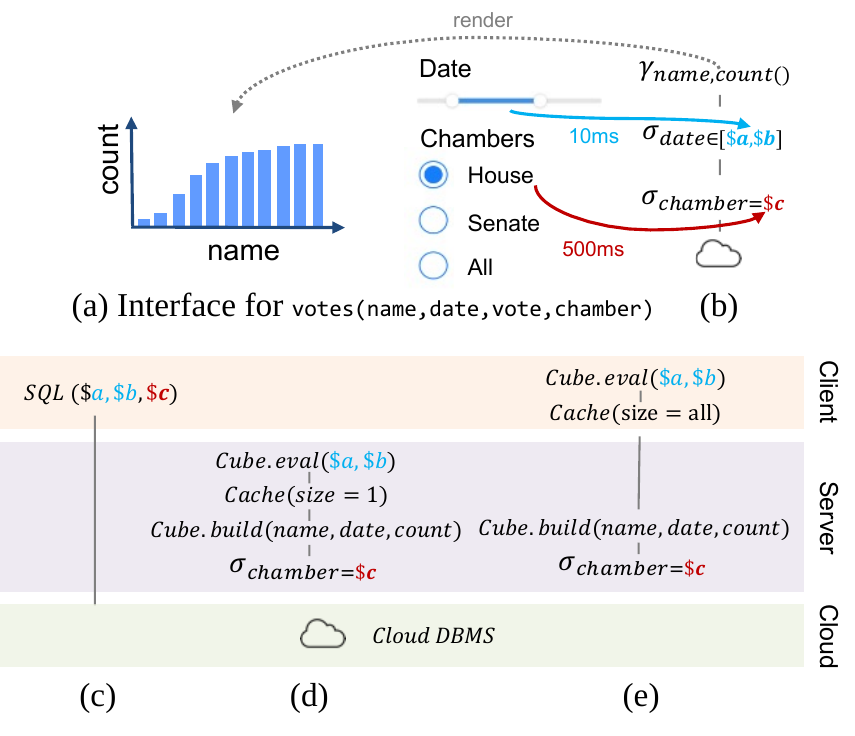}
    \caption{(a) Interface to analyze vote counts by congressional member.  (b) Interactions bind to {\it choices} in a logical query plan.  Three candidate execution plans that (c) send queries to the cloud DBMS, (d) pre-compute and evaluate interactions using cubes on the server, and (e) caches cubes on the client. }
    \label{fig:pvdarch}
\end{figure}

\begin{example}

To illustrate, \Cref{fig:pvdarch}(a) visualizes vote counts by congressional member, with user-controlled parameters for chamber and date range. 
Logically, the visualization corresponds to a simple SPA query: filter by chamber and date range, then group by member and count votes. 
Interactions bind concrete values to parameters in this plan, and designers may specify latency constraints (e.g., sliders must update within 20ms; changing the chamber may take up to 500ms). PVD uses this representation to reason about feasibility and to generate an execution plan that sends updated results to the view's \code{render()} methods.


\Cref{fig:pvdarch}(c–e) shows three candidate plans.
In (c), the client issues SQL queries directly to the cloud DBMS---this is simple but slow.
In (d), the server retrieves the table from the cloud DBMS, filters the table, builds a cube over \texttt{name} and \texttt{date}, and caches it.   This accelerates date changes but needs to be recomputed and recached if the chamber changes.  
In (e), the client caches cubes for all parameter assignments.  This is fastest, but requires substantial client memory.
The ``right'' plan depends on query structure, client and server resources, network properties, and latency expectations. PVD’s role is to automatically explore this space; the designer specifies only the interface and its constraints.


\end{example}


At first glance, PVD should not be possible.  PDD is already considered intractable due to challenges of cost estimation and a combinatorial search space, and solutions only attempt to reduce expected workload latency.   PVD appears to be strictly harder because it guarantees wall-clock latency for interactions.
Surprisingly, we find that these millisecond-to-second latency constraints actually simplify the search because they eliminate the hardest cases in query optimization: joins with large or unpredictable fan-out. If a join cannot guarantee bounded output size~\cite{armbrust2009scads}, it cannot satisfy the latency constraints and is pruned. This dramatically reduces the search space, making PVD feasible despite its stronger guarantees.


\subsubsection{An Instance of Physical Visualization Design}

Our system, Jade~\cite{chen2025physical}, is the first instantiation of PVD that focuses on choosing and placing data structures for interactive visualization interfaces. Because modern visualizations rely on bespoke data structures (e.g., cubes, sketches, indexes), Jade is designed to support an extensible library of custom data structures.  

Jade introduces a compact interface representation that extends logical query plans with {\it Choice} operators. Choices parameterize literals, expressions, or entire subplans.  For example, a dropdown may {\it Choose} between two subqueries, or a range slider may bind Choices for lower and upper bounds of a predicate. A single plan thus represents a combinatorial set of concrete plans, much like how a context-free grammar represents a set of valid strings. These plans provide the structure for analysis and rewrite rules.



To integrate bespoke data structures into rule-based optimization, Jade models them as a typed pair of sub-operators, that mirror traditional pipeline breakers: \code{build()} converts a table into a data structure and \code{eval()} consumes a data structure and bindings and returns a table.   This makes it possible to freely place caching and network operators to control fine-grained data placement.  

\begin{example}
\Cref{fig:pvdarch}(d) uses \code{Cube.build()} to compute a byte array that encodes a dense cube, the \code{Cache()} operator to reuse the cube when \code{\$a} or \code{\$b} change, and \code{Cube.eval(2001, 2020)} uses the cube to compute the vote distribution between 2001 and 2020.  Data structures also estimate their cost and size based on statistics, and specify a \code{match()} method to identify subplans they can replace. 
\end{example}

\begin{figure}
    \centering
    \includegraphics[width=\linewidth]{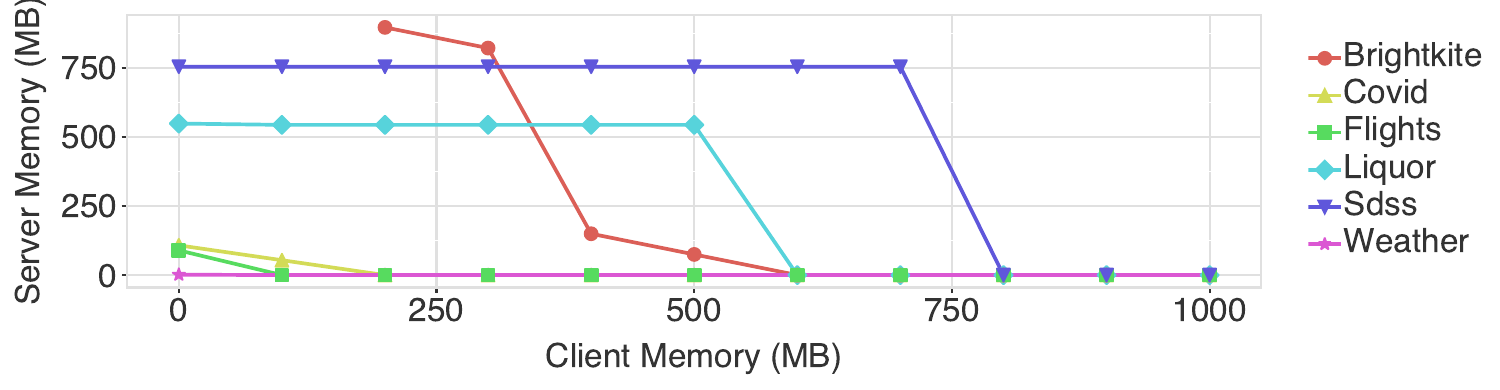}
    \caption{Pareto curve for the client and server resources needed to meet latency expectations for 6 interactive interfaces.   }
    \label{fig:pvd-pareto}
\end{figure}

\Cref{fig:pvd-pareto} illustrates these trade-offs across six benchmark interfaces that vary in complexity and interaction patterns. Latency constraints are set to $20{ms}$ for continuous interactions (e.g., brush) and $200{ms}$ for discrete interactions (e.g., click). Simple interfaces (e.g., Covid, Weather) require few resources; more complex ones (e.g., Brightkite, with many coordinated interactions) exhibit rich Pareto frontiers. As data size, network conditions, and latency constraints shift, these frontiers change, demonstrating that no single optimization is robust across settings. An automated optimizer is essential for navigating this high-dimensional space quickly.

Jade shows that PVD can analyze an interface representation and automatically synthesize efficient execution strategies across heterogeneous deployments. It also demonstrates that design independence is tractable in practice: a simple rule-based optimizer can cover a surprising range of realistic interfaces once guided by a principled interface model.



\subsection{What Next?}

PVD highlights a broad challenge in HDI systems: modern data interfaces sit atop an increasingly fragmented ecosystem of databases, execution engines, and hardware accelerators. Just as physical database design helps architects navigate diverse indexing and storage choices, interface designers now require tools that can reason across heterogeneous backends, interaction patterns, and resource constraints.
PVD addresses this challenge by matching two formal models:
\begin{itemize}[leftmargin=*,itemsep=0em]
    \item An {\it Interface Model} that captures the interface's structure---its data flow, interactions, perceptual constraints, and latency requirements, and
    \item A {\it System Model} that captures client-server architecture, execution strategies, diverse data structures, caching and placement options, and performance guarantees. 
\end{itemize}
Given these models, PVD can systematically search for feasible execution plans, or determine that none exist—providing designers with immediate feedback without blocking on an engineering team.

Jade instantiates this idea for the static relational setting by modeling interface behavior through Choice-augmented plans and modeling system capabilities through rule-based rewrites over data-structure operators. This demonstrates that, given these models, it is tractable to automatically synthesize efficient and useful end-to-end system designs.

Generalizing this approach requires richer and more expressive models on both sides.
On the interface side, models must eventually capture coordinated multi-view behavior, layout and rendering costs, streaming updates, and user-perception constraints—not just query-level semantics.
On the system side, the model must evolve to represent heterogeneous execution backends (e.g., cloud warehouses, stream processors, in-browser engines), consistency choices, fine-grained caching, and incremental or learned components.

Bridging these models would support principled synthesis of complete HDI architectures, reduce reliance on bespoke engineering, and empower interface designers to rapidly iterate on designs. Over time, such model-driven synthesis could provide a foundation for data interface design that separates high-level intent from the details of efficient execution.  

\section{From Systems to New Interfaces}\label{s:system}

\begin{figure}
    \centering
    \includegraphics[width=0.65\linewidth]{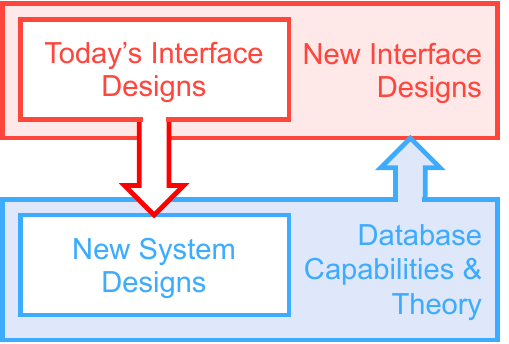}
    \caption{ Database capabilities and theory go well beyond the system designs and optimizations developed in response to {\it today's} interface designs, and can motivate new types of interface and interaction designs, as part of a virtuous cycle.  }
    \label{fig:s2i}
\end{figure}

The previous section showed how formal models of an {\it existing interface} can guide the design of new system optimizations. Yet this direction of influence has an important limitation: interfaces often reflect what is easy to build with current technologies rather than what best supports human analysis. This creates a subtle ``streetlight effect,'' where both interface and system co-evolve around local optima defined by present-day implementation constraints.

To break out of this local optimum, we observe that visualization is not a fixed application domain like genomics or finance. \textbf{Visualization is workload-agnostic: it aims to support human reasoning over whatever the user's data, tasks, and analytical goals are.} Its design space inherits the full complexity of data models, uncertainty, interaction patterns, and cognitive demands. In this sense, visualization is as close to data management as to end-user applications: it is a general-purpose substrate that must adapt to whatever data and analytical questions arise.

This section inverts the design flow (Figure~\ref{fig:s2i}). 
Instead of asking {\it ``How can systems support current interfaces?''}, we ask {\it ``What new interface capabilities become possible when systems provide stronger abstractions, semantics, or guarantees?''}
These, in turn, impose new requirements that motivate further system innovations, creating a virtuous cycle.

We illustrate this with four projects from our lab that show how distinct properties of data systems expand the interface design space.
Literature on {\it data cube analytics} motivates compositional interactions that operate over entire data-processing pipelines.
{\it Query workloads} can be modeled as grammars, providing compact representations from which interfaces can be synthesized.
{\it Relational semantics} enable multi-table analytics and expose new forms of interaction unavailable under single-table assumptions.
And {\it relational theory} itself allows us to extend visualization from mappings over a single table to designs that reflect the full relational structure of a database.
Together, these properties demonstrate how system abstractions can drive new forms of interaction and visualization that current tools cannot express.


\subsection{From Query Workloads to New Interactions}

Our first project explores how structure in workloads can motivate novel interaction designs.
Interaction in visualization systems involves both manipulating visual elements and applying the corresponding data transformations required to update the view. Most existing systems expose a small set of well-understood interaction techniques---creating new views, manipulating a view, and coordinating multiple views---that map cleanly onto modifications of SQL \code{SELECT} and \code{WHERE} clauses.

\begin{figure}
    \centering
    \includegraphics[width=1\linewidth]{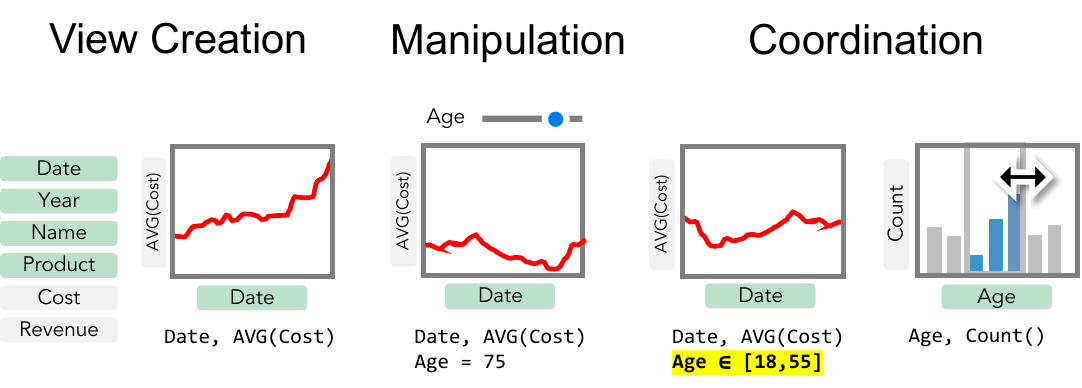}
    \caption{Existing interactions are focused on creating new views from data, changing a view by manipulate controls, and coordinating the updates in one view through interactions of a separate view. These interactions are largely limited to modifying SQL \code{SELECT} and \code{WHERE} clauses.   }
    \label{fig:interactiontaxonomy}
\end{figure}

\Cref{fig:interactiontaxonomy} illustrates these three categories.
\textit{Create} interactions generate new views (e.g., dragging \code{Date} to the x-axis constructs a group-by query).
\textit{Manipulate} interactions modify an existing view, such as filtering or changing an aggregation.
\textit{Coordination} interactions, such as brushing and linking, propagate predicates across views.
These interactions are simple to implement because they correspond directly to literal substitutions inside a single-SPA query.

While popular, this library largely reflects what {\it today's systems} can execute efficiently, rather than the full range of analytical operations users perform. For example, what if the user wishes to  compare data slices, e.g.,  {\it ``the average cost over time for 75-year-olds against the overall population?''} Such comparisons are commonplace in OLAP and cube-based analytics~\cite{gray1997data,sarawagi2000i3} but have no counterpart in modern visualization systems because they require reasoning about the semantics and schemas of multiple derived views, not just editing a predicate in a single query.

\begin{figure}
    \centering
    \includegraphics[width=1\linewidth]{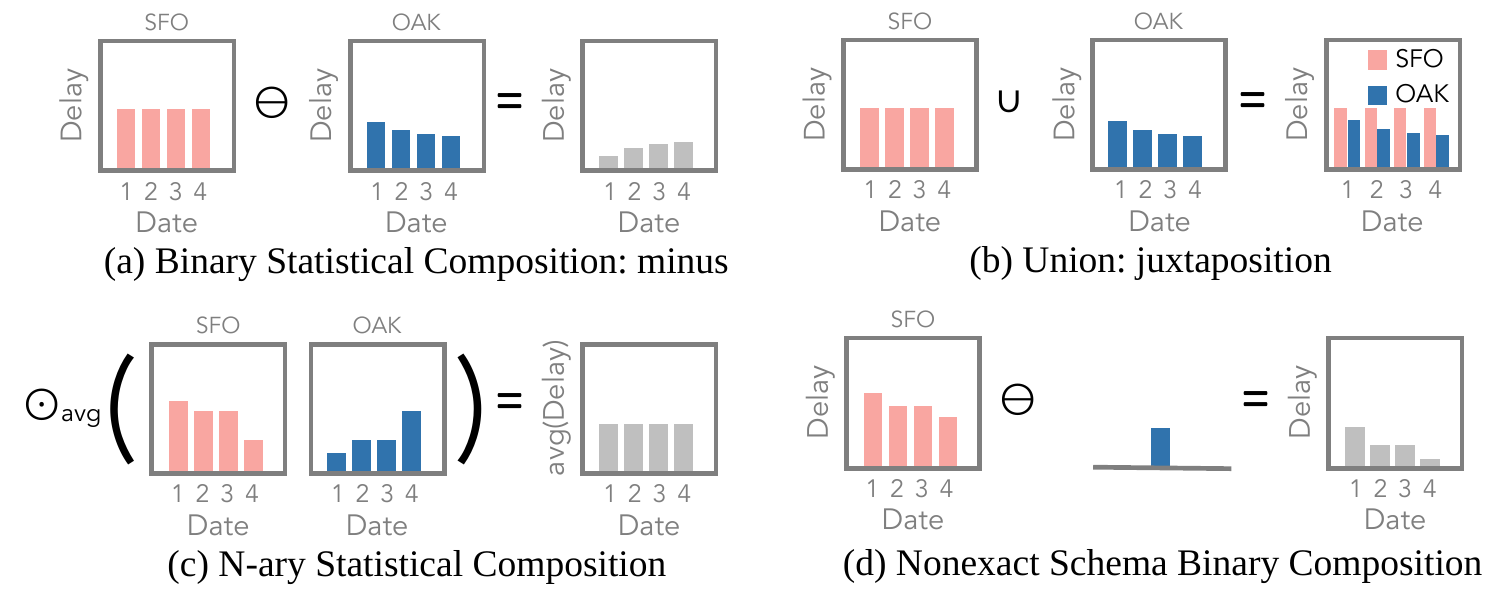}
    \caption{Example view composition operations that (a) subtract one view from another, (b) union two views to render their data together, (c) compute the average based on their join keys, and (d) subtract a scalar from a bar chart. 
 }
    \label{fig:vcaops}
\end{figure}

\begin{figure}
    \centering
    \includegraphics[width=.75\linewidth]{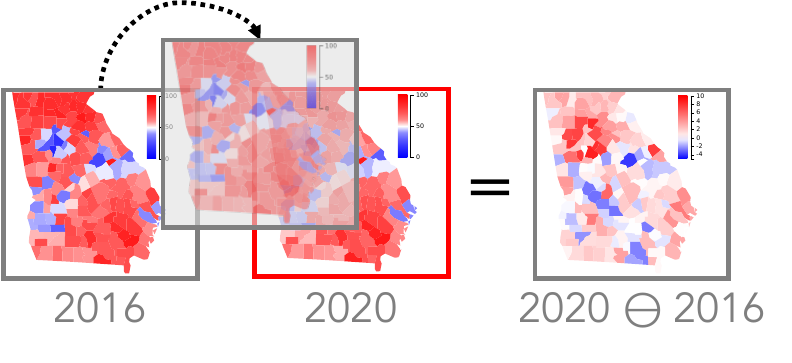}
    \caption{Dragging the 2016 election results over 2020 computes the difference by county.}
    \label{fig:vcamap}
\end{figure}

This gap motivated our View Composition Algebra work (VCA)~\cite{wu2022view}, which treats views as first-class algebraic objects that can be composed, joined, merged, or differenced.  Since views are the output of query pipelines, VCA operates over entire data-processing pipelines; this is in contrast to traditional interactions that modify SQL literals.
\Cref{fig:vcamap} demonstrates VCA in practice: the user drags the 2016 election results onto the 2020 results to compute their difference and visualize the result as a map---with no manual data wrangling.
\Cref{fig:vcaops} shows 4 illustrative VCA operators: (a) element-wise subtract each mark in one view with its match(es) in the other, (b) unioning the records in two views together, (c) computing the average of each date's delays across all cities, and (d) implicitly joining the underlying tables to subtract the blue bar from every date.

Ideally, a user can freely compare any combination of data visualized in an interface.  However, supporting such {\it ad hoc} comparisons in visual analysis tools is tricky because the specific data that will be visualized is often not known ahead of time and not all data is directly comparable.
Thus, ad hoc comparisons require the system to dynamically assess whether a combination of views is valid to compose. From a data management perspective, this reduces to classic data integration problems:
{\it Are two views comparable?} That is, do their schemas admit a semantically valid mapping (e.g., \texttt{cost} $\sim$ \texttt{price})?
If so, {\it which tuples should be matched?} This is entity resolution.
Data integration remains an open problem, and thus, advances in schema matching, type inference, and entity resolution directly expand what comparison interactions can support.

\smallskip
Comparison is just one example of interaction techniques made possible by data-management concepts. Another is explanatory interactions, which surface predicates that explain surprising visual patterns~\cite{wu2013scorpion}. But enabling explanations inside interactive systems required nearly a decade of systems innovations from our lab---to support fast lineage capture~\cite{Psallidas2018SmokeFL,Psallidas2018ProvenanceFI,Psallidas2018DemonstrationOS} and scalable lineage evaluation~\cite{mohammed2025fade}---that combines with prior work on fast explanation discovery algorithms~\cite{abuzaid2018diff,roy2015explaining,gathani2020debugging}. 
These systems-level innovations need to be in place before such interactions are feasible at the latency expected from modern interfaces.

\subsection{From Query Workloads to New Interfaces}

Our second project shows how the compositional structure in query workloads can be used to support automatic interface synthesis to go beyond today's template-based approaches to interface design.

While using workload characteristics to inform interaction design is powerful, many modern interfaces still require substantial manual engineering. Tools such as Retool or Hex aim to avoid this by translating SQL queries into dashboards; however, their interaction models are limited, and developers must handcraft most interface logic (\Cref{ss:pvd}).  What if it were possible to {\it derive an entire interface} directly from a workload in a way that preserves its full expressive power?

To do so, we need a representation between a list of queries and a full interface implementation: abstract enough to generalize, but structured enough to generate concrete interface components. Let us step back to reflect on what a data interface really is: a set of interactive controls that predictably modify and issue queries, whose results are rendered as visualizations in the interface. In a sense, an interface is an interactive representation of its set of expressible queries: a slider expresses a numeric literal in a filter predicate while a command-line interface expresses the entire SQL grammar.

We formalized this idea as the Data Interface Grammar (DIG)~\cite{dig}. DIG models interfaces as grammars whose non-terminals correspond to interface choices and whose productions encode schema-aware SQL structure. DIG extends Parsing Expression Grammars (PEGs)~\cite{peg} with SQL- and schema-specific annotations that constrain allowable expressions. A valid interface is a visual representation that expresses this grammar---user interactions resolve non-terminals to produce a concrete parse tree, which corresponds to a valid query.


\begin{figure}
    \centering
    \includegraphics[width=\linewidth]{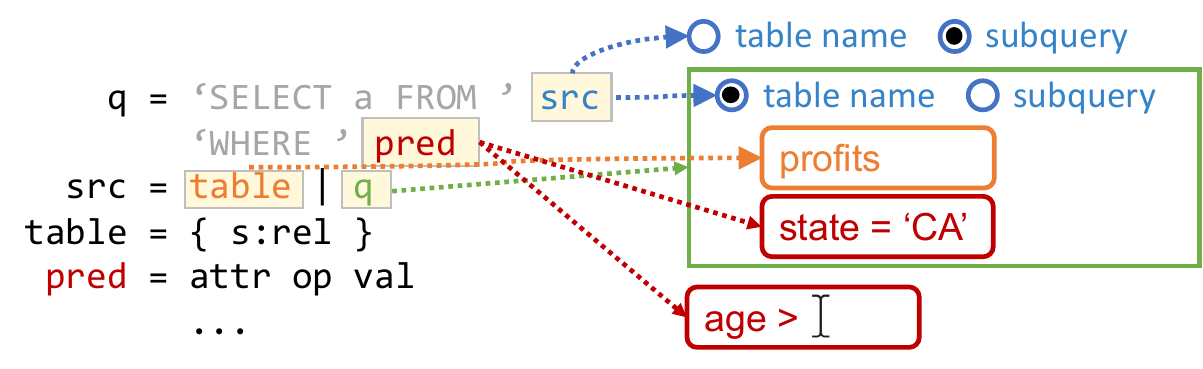}
    \caption{A simple query building grammar and mapping to interface.}
    \label{fig:dig}
\end{figure}

\begin{example}
  \Cref{fig:dig} illustrates an example grammar for a query builder.  
  The query consists of a fixed string template, but the \code{src} and \code{pred} are chosen by the user. \code{src} maps to a radio button where selecting the first option shows a text box to select a valid relation name (enforced by the \code{table} production rule), while the second option recursively nests the same interface. 
  \code{pred} maps to a text box that parses the input against the grammar to ensure it is valid.   Alternatively, it could map to three widgets for the attribute, operator, and value.
\end{example}

This representation offers several benefits. 
First, applications don't assemble raw SQL strings and every query conforms to the grammar.
Second, grammar rewrites offer a simple mechanism to refactor and explore alternative interface designs.
Third, DIG aligns naturally with existing data engineering practices.  For instance, DBT pipelines already structure analytics as DAGs of SQL views, which DIG can express and directly map to interactive dashboards.  
Fourth, DIG shares structural similarities with the plan-based representations used in PVD (\Cref{ss:pvd}), making it amenable to automatic optimization.
Finally, standard techniques such as grammar induction can infer DIG grammars from query traces, enabling interface synthesis---an approach we explored in a line of work called Precision Interfaces~\cite{pi,pi2,pinotebook,nl2interface}.

DIG is also useful when using large language models (LLMs).   Natural language queries often contain unresolved ambiguities in intent, schema interpretation, or operator choice~\cite{nlambig}. Rather than forcing the system to commit to a single query, the LLM can simply represent ambiguity as non-terminals and return an interface whose interactions navigate the space of ambiguities. This idea underlies NL2Interface~\cite{nl2interface}, where LLMs generate a grammar expressing possible interpretations, and the user resolves them through interaction.

\subsection{From Single Table to Multi-Table Analytics}\label{ss:semiring}

Our third project examines how interactive analytics can be extended from single- to multi-tables by leveraging database theory in factorized query execution. 

Today’s interaction designs are also limited by what today’s systems efficiently support. For instance, most visualization systems are designed under the assumption that the data resides in a single table or view. This assumption enables optimizations like data cubes, which exploit the distributive properties of common aggregation functions to materialize partial aggregates and combine them efficiently to support fast drill-down, roll-up, and slicing operations. Despite join execution being a major pillar of data management research, cube pre-computation techniques do not extend to joins, and users cannot interactively specify join conditions or change source relations.

Instead, data engineers predefine denormalized views that join relevant tables~\cite{denorm,denorm2}, typically managed through so-called \textit{semantic layers}. This incurs substantial storage overhead if the joins are fully materialized, or introduces latency if executed per interaction.

Although not applied to visualization systems, the theoretical foundation to support interactive analytics over joins is well established.   Probabilistic graphical models are akin to count-based analytics over a join graph. As early as the 1980s, the machine learning literature developed calibrated junction trees~\cite{shafer_shenoy1988,pearl1988probabilistic} as a partial aggregation data structure to support fast, exact marginal computations.  We developed the Calibrated Junction Hypertree~\cite{Huang2023LightweightMF} to extend these ideas to any semiring aggregate and designed a set of reuse techniques to support OLAP and join operations (add, remove, change joins) in interactive latencies.

This system serves as the basis for designing join-oriented interactions and interfaces. Supporting multi-table semantics at the system layer also revealed subtle correctness errors that current interfaces silently inherit.  Aggregations over 1-N or N-M joins can unintentionally duplicate data in ways that corrupt statistics and lead to incorrect visualization contents~\cite{Huang2023AggregationCE,hyde2024measures}.   
For example, ad-attribution pipelines often join users to impressions, causing user-level metrics to be replicated across many impressions and leading to inflated aggregates.
A principled solution requires the system to track record multiplicities and for the interface to surface or infer the user’s intended aggregation semantics, so that each input entity contributes exactly once.

\subsection{From Single Table to Database Visualization}

While the previous subsection showed how to perform interactive analytics over multi-table schemas, it focused on the data layer: {\it visualization design still assumes a single-table input}. Nearly all graphical grammars—rooted in Bertin’s data–visual mappings and formalized in the grammar of graphics \cite{Wilkinson2001nViZnA,Wickham2010ALG,stolte2002polaris,Satyanarayan2018VegaLiteAG,wilkinson2012grammar}—treat a visualization as a mapping from a table: tuples become marks, attributes become visual channels, and the schema determines the structure of the view.

This assumption clashes directly with relational theory. Real databases represent information across multiple relations with keys and constraints that preserve meaning. Forcing a join collapses this structure into a denormalized table whose tuple multiplicities and foreign-key relationships rarely reflect the original schema. As a result, flattening multi-table data can distort semantics: keys become duplicated, relationship structure is lost, and multiplicity artifacts can influence aggregates or visual patterns.

Consequently, existing visualization grammars do not guarantee semantically faithful representations of multi-table data. What is needed is a theory and system design that can visualize multi-table databases without first reducing them to a single table.

\begin{figure}
    \centering
    \includegraphics[width=1\linewidth]{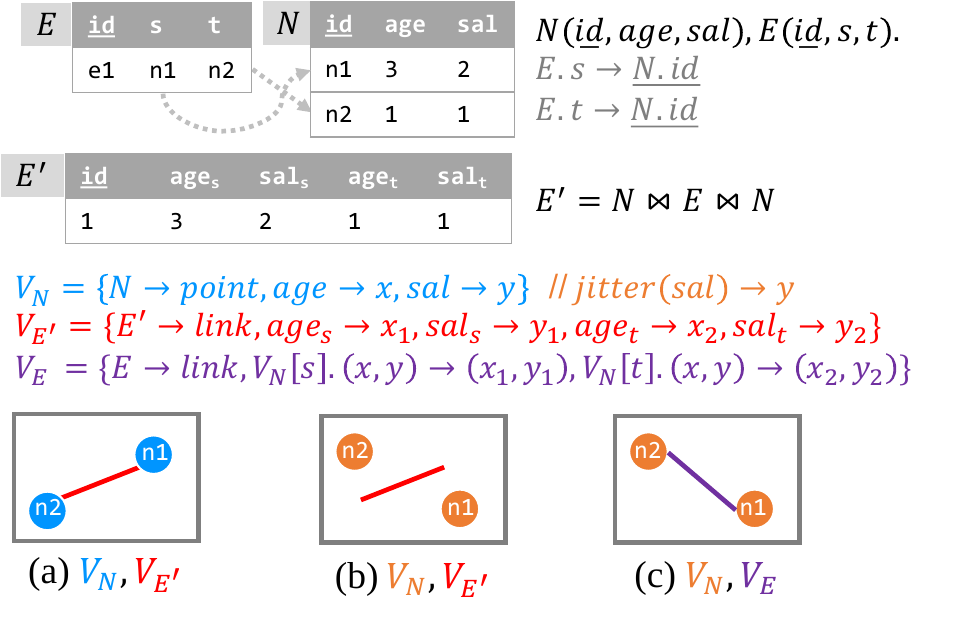}
    \caption{Node link visualizations of the edges and nodes tables.    (a) \red{$V_{E'}$} only appears to connect the points \blue{$V_N$}, but (b) become inconsistent if \orange{$V_N$} changes (e.g., adding \orange{jitter} to the y layout). (c) \purple{$V_E$} preserves the foreign key relationships by referencing the mark positions (e.g., \purple{$V_{N}[s].(x,y)$}).}
    \label{fig:nodelink}
\end{figure}

\subsubsection{Illustrative Example: Node-Link Layouts}

\Cref{fig:nodelink}(a) visualizes a nodes table $N(id, age, sal)$ as a scatter plot \blue{$V_N$}, and overlays edges from the table $E(id,s,t)$ by first constructing the denormalized join $\red{E'} = N \Join E \Join N$ and mapping \red{$E'$} to lines.
The edges, however, only appear to connect the nodes. If the user applies \orange{jitter}, the node positions change but the edges do not (\Cref{fig:nodelink}(b)).  This inconsistency violates the underlying relationship because the visualization lacks a representation of the data's foreign-key constraint.

This limitation affects both how multi-table visualizations are {\it specified} (as above) and which {\it designs} are even expressible. Designs such as parallel coordinates, Sankey diagrams, or space-filling layouts internally encode multi-table structure but do so outside any visualization grammar, and instead rely on custom code or ad-hoc conventions to preserve relational semantics. What is needed is a principled extension from ``Table Visualization'' to Database Visualization \cite{dvlicdt,dvlarxiv}.

\subsubsection{Database Visualization}

Our core observation is that a visualization does not merely map tuples to marks; it also maps {\it database constraints to visual representations of these constraints}. Tables and schemas map to mark types; attribute domains map to axes and scales; and keys and foreign keys map to visual relationships that keep marks consistent with the underlying data model.

From this perspective, extending the relational model from a single table to a database corresponds to extending visualization designs to preserve multi-table semantics. This can be achieved by referencing foreign-key relationships directly in the visualization specification (as shown below), rendering them explicitly as marks, or encoding them implicitly via spatial arrangements such as containment or alignment.

\begin{example}\label{ex:nodelink2}
Database visualization theory preserves the foreign key relationships between $N$ and $E$ in \Cref{fig:nodelink}(c) by defining \purple{$V_E$} to reference the mark positions of the circles that render the source and target nodes (e.g., \purple{$V_{N}[s].(x,y)$}).  If $n_1$ moves, its incident edges always remain visually consistent.
\end{example}

\begin{figure}
    \centering
    \includegraphics[width=\linewidth]{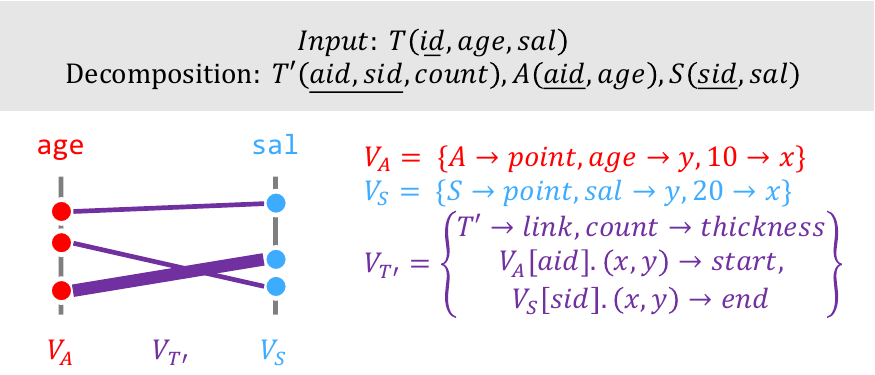}
    \caption{Under data visualization theory, parallel coordinates are a consequence of data modeling decisions to emphasize unique values of each attribute and the number of relationships between them.}
    \label{fig:dvl-parallel}
\end{figure}

Database visualization has theoretical and practical implications. Theoretically, it unifies a diverse set of visualization designs---including trees, parallel coordinates, flow diagrams, faceting, and nested visualizations---and shows that they are byproducts of data decomposition and simple visual mappings rather than bespoke visual forms. 

\begin{example}
    \Cref{fig:dvl-parallel} illustrates how parallel coordinates emerge naturally from normalizing age and salary to render each as points ($\red{V_A}$,
    $\blue{V_S}$),
    aggregating $T$ by \code{aid, sid}, and drawing lines 
    $\purple{V_{T'}}$ between points whose thickness reflects the count.  
\end{example}

This perspective reframes denormalization as one in a combinatorial space of possible database decompositions, which each lead to different valid visual representations. Visual analysis now becomes a joint problem of data modeling and visual encoding, which mirrors how data models guide application development.

A theory of database visualization enables the automatic generation of expressive and semantically correct visualizations for arbitrary structured data. This opens new possibilities: administrators can visualize entire databases; users can explore personal data holistically; data scientists can track intermediates in a workflow; and analysts can browse large repositories with awareness of schema-level relationships and constraints.

\smallskip
These projects demonstrate how systems-level innovations expand the space of interactions and visual designs. When systems expose algebraic structure, query grammars, multi-table semantics, and relational constraints, interfaces can support analyses that are currently impossible or prohibitively labor-intensive. Rather than treating interface design as bounded by the capabilities of today's systems, we can instead view systems and interfaces as mutually reinforcing layers that evolve virtuously together, each enabling new forms of human reasoning over data.

\section{What About AI?}

Large language models (LLMs) are sequential neural networks pre-trained on web-scale text corpora~\cite{vaswani2017attention}. They have demonstrated impressive capabilities in language understanding, reasoning, domain knowledge, programming~\cite{yang2024sweagent}, and data tasks~\cite{shankar2024docetl,tan2024large,demiralp2024making}. 
Abstractly, an LLM acts as a function $f(any)\to any$: any task that can be serialized as text can be used as input and interpreted as output. 
Yet, this apparent generality hides a deep uncertainty. User goals may be under-specified, prompt descriptions imprecise, and the model’s outputs are probabilistically sampled from large, unstructured distributions. As a result, guarantees of safety, correctness, and performance remain elusive.


Despite these limitations, LLMs are already being used to augment or replace parts of data analysis and visualization pipeline, including data modeling~\cite{kotelnikov2023tabddpm,zhang2023mixed}, SQL generation~\cite{satriani2025logical}, dashboard creation~\cite{dibia2023lida,wang2025data,chen2024viseval}, exploratory analysis~\cite{dibia2023lida}, and even product ideation~\cite{productstrategy}. This breadth naturally raises the question: {\it if LLMs can seemingly do everything, what role remains for HDI systems research?}

To explore this question through the lens of interfaces, consider the rapidly growing practice of vibe coding~\cite{vibecoding}: generating functional artifacts through natural-language interaction with an LLM or agent. Tools such as Cursor, Replit, Lovable, and Convex Chef have gained rapid adoption by lowering the barrier to implementing data workflows. An Anthropic blog post reports using Claude Code to accelerate infrastructure, product design, growth, and data science development~\cite{claudecode}.   They state that  {\it ``despite knowing `very little JavaScript and TypeScript,' the team uses Claude Code to build entire React applications for visualizing RL model performance and training data.}  
These examples suggest that LLMs can now synthesize non-trivial user interfaces for complex data tasks, dramatically reducing the friction of manual UI development.

However, these interfaces still depend on robust backend systems. As Section~\ref{s:interface} illustrated, even achieving interactive latencies is technically demanding, let alone ensuring consistency, reliability, or safety. LLMs lack the performance models, hardware awareness, and correctness invariants needed to make systems-level trade-offs: storage layout, physical operator choices, concurrency, and cache reuse. Empirical studies confirm these gaps. LLM accuracy collapses as task complexity grows~\cite{lin2025zebralogic,illusion}; models fail to apply even simple compiler rewrites~\cite{fang2024llmpeep}; and they often reason shallowly about programs, with poor generalization~\cite{cummins2023llmcomp}. LLM-generated code can orchestrate components, but guarantees must come from the underlying systems.

This separation of concerns mirrors other domains. Infrastructure-as-code tools expose constrained declarative APIs that Copilot can reliably work with; compilers expose structured IRs like LLVM; browsers expose composable DOM abstractions. In each case, LLMs depend on system-layer primitives that are explicitly designed for predictability, scope, and performance.

For HDI, the implication is clear: LLMs do not replace systems---they are clients. The role of HDI systems research is to define the abstractions that agentic software relies on for reliably, performance, scalability, and safely over data. 
These abstractions must help meet user-expectations, such as strict latency bounds for interactivity, mechanisms to enforce correctness, and provenance-based constraints to avoid silent failures. Incremental improvements are rarely sufficient, and if a system falls short of these thresholds, it is often unusable. HDI systems researchers have an opportunity to build the substrate that makes agentic software design for HDI viable at scale.


\section{Conclusions}

Human–data interaction (HDI) is collectively shaped by the user's analytical needs and the capabilities and constraints of the underlying systems. The past decade of research reviewed here illustrates how progress in HDI comes from co-design of systems and interfaces, rather than from advances in either layer alone. This paper has argued that HDI is not merely a user-interface concern ``thrown over the fence'' to systems experts, but a systems challenge that must grapple with latency, semantics, interactivity, uncertainty, and perception. Historically, interface designs and visualization formalisms have motivated new system optimizations; equally, database theory, query expressiveness, and system-level semantics can inspire new visualization theories and interaction designs.

Finally, AI is not a panacea. LLMs depend on the guarantees that systems provide in order to safely generate and execute code that manipulates state. Rather than displacing systems research, AI increases the need for principled abstractions that ensure correctness, interactivity, and trust.  
The past decade of work demonstrates that when systems and interfaces co-evolve, HDI can support richer, faster, and more trustworthy forms of human reasoning---an agenda that remains as essential now as ever.

\section{Acknowledgements}
We thank Jiaxiang Liu,  Charlie Summers, Haonan Wang,  Lampros Flokas, Fotis Psallidas, Thibault Sellam, Qianrui Zhang, Haoci Zhang, Viraj Rai, Yifan Wu, Larry Xu, Daniel Alabi, Xiaolan Wang, Niranjan Kamat, Lilong Jiang, Gabriel Ryan, Ziyun Wei, Zhengjie Miao, Marianne Procopio, Ab Mosca, Lana Ramjit, Jacob Fisher, Zhaoning Kong, Jeff Tao for their contributions to the research lab, ideas, and the cited projects.  We are grateful to Remco Chang, Joe Hellerstein, Arnab Nandi, Yunhai Wang, Subrata Mitra, Ravi Netravali, Jiannan Wang, Alexandra Meliou, Leilani Battle, Sam Madden, Carlos Scheidegger, Aditya Parameswaran, Javad Ghaderi, Dan Rubenstein for collaborations and formative discussions over the years.

This research received funding from the National Science Foundation grants (NSF  1527765, 1564049, 1845638, 1740305, 2008295, 2106197, 2103794, 2312991) as well as corporate support from Amazon, Google, Adobe, CAIT, and Intellect Design. The views and conclusions presented here are those of the
authors and should not be interpreted as representing the official positions of the funding organizations.

\bibliographystyle{elsarticle-num} 
\bibliography{main}

\end{document}